\documentclass{elsart}

\usepackage[dvips]{graphicx}

\usepackage{amssymb}

\begin{document}

\begin{frontmatter}



\title{Measurement of Single and Double Spin-Flip Probabilities 
in Inelastic Deuteron Scattering on \nuc{12}{C} at 270 MeV} 

\author[Riken,email]{Y. Satou},
\author[Riken,present]{S. Ishida},
\author[Tokyo_u]{H. Sakai},
\author[Saitama_u]{H. Okamura},
\author[Riken]{H. Otsu}, 
\author[Riken]{N. Sakamoto}, 
\author[Saitama_u]{T. Uesaka}, 
\author[Riken]{T. Wakasa}, 
\author[Tokyo_u]{T. Ohnishi}, 
\author[Tokyo_u]{K. Sekiguchi}, 
\author[Tokyo_u]{K. Yako}, 
\author[Riken]{T. Ichihara}, 
\author[TIT]{T. Niizeki}, 
\author[Riken]{K. S. Itoh}, 
\author[JAERI]{N. Nishimori} 
\address[Riken]{The Institute of Physical and Chemical Research (RIKEN), 
Wako, Saitama 351-0198, Japan}
\address[Tokyo_u]{Department of Physics, University of Tokyo, Bunkyo, Tokyo 
113-0033, Japan}
\address[Saitama_u]{Department of Physics, Saitama University, Urawa, Saitama 
338-8570, Japan}
\address[TIT]{Department of Physics, Tokyo Institute of Technology, 
Oh-okayama, Tokyo 152-8551, Japan}
\address[JAERI]{Japan Atomic Energy Research Institute (JAERI), 
Ibaraki, 319-1195, Japan}
\thanks[email]{E-mail : ysatou@rarfaxp.riken.go.jp}
\thanks[present]{Present address: Soei International Patent Firm, Tokyo 
104-0031, Japan.}

\begin{abstract}
The deuteron single and double spin-flip probabilities, 
$S_1$ and $S_2$, 
have been measured 
for the \nuc{12}{C}$(\pol{d},\pol{d}')$ reaction at E$_d$ = 270 MeV 
for an excitation energy range between 4 and 24 MeV 
and a scattering angular range between $\Theta_{\rm lab}$= 2.5 
and 7.5$^{\circ}$. 
The extracted $S_1$ exhibits characteristic values 
depending on the structure of the excited state. 
The $S_2$ is close to zero 
over the measured excitation energy range. 
The SFP angular distribution data 
for the 2$^+$ (4.44 MeV) and 1$^+$ (12.71 MeV) states 
are well described by the microscopic DWIA calculations. 
\end{abstract}

%
\end{frontmatter}

Studies of polarization phenomena in nucleon induced inelastic scattering 
at intermediate energies (100-1000 MeV) 
have been one of the active fields of research in nuclear physics. 
Not only giving an insight into the reaction mechanisms, 
the measurement of spin-flip probability (SFP), $S_{nn}$ in particular, 
provided useful information on the spin-flip modes 
of nuclear excitation~\cite{Baker97}. 
At the intermediate energy region 
the spin transitions are dominated by the isovector ones 
due to the overwhelmingly large vector-isovector 
component of the effective interaction. 
Therefore, 
except for a few isolated states~\cite{Crawley89}, 
there is still limited knowledge on the isoscalar spin excitations 
by the nucleon induced reactions as well as by other conventional probes. 

The polarization transfer measurement in inelastic deuteron scattering 
should be one of the essential probes of the isoscalar spin excitations. 
Since deuterons have an isospin quantum number of $T$=0, 
the reaction is selective of isoscalar transitions. 
The spin $S$=1 nature of the deuteron 
allows spin transfer to the target. 
The measurement of deuteron SFP 
should provide a means to disentangle between spin and non-spin excitations, 
and one may expect 
that a similar polarization technique used to extract information 
on the spin-dependent structure from nucleon scattering 
may also be exploited for deuteron case. 

The deuteron single and double SFPs, 
defined as fractions of deuteron undergoing spin-flip 
by one and two units along an axis normal to the reaction plane, 
are expressed in terms of polarization observables by the relations: 
\begin{eqnarray}
S_1&=&\frac19(4-P^{y'y'}-A_{yy}-2K_{yy}^{y'y'}), \\
S_2&=&\frac1{18}(4+2P^{y'y'}+2A_{yy}-9K_{y}^{y'}+K_{yy}^{y'y'}). 
\end{eqnarray}
The quantities $A$, $P$ and $K$ 
refer to the analyzing power, polarizing power 
and polarization transfer coefficient, 
one (two) indices stand for the vector (tensor) polarization, 
and lower (upper) ones the incident (outgoing) beam. 
The determination of $S_1$ and $S_2$ thus requires 
vector and tensor polarized beams and vector and tensor polarimeters. 
The SFPs, however, represent the projectile spin-flip 
and do not necessarily indicate the target spin-flip~\cite{Tsuzuki91}. 
One thus needs to rely on a certain reaction model to interpret the SFP data 
in terms of the effective interaction and the target structure. 

The earliest measurement of $S_1$ 
utilized the $(d,d'\gamma)$ method 
for states in \nuc{12}{C}~\cite{Sishida93}. 
The correlation measurements of the scattered deuterons 
and the emitted $\gamma$-rays 
could yield $S_1$ 
for states with known $\gamma$ decay branch to the $0^+$ ground state. 
A direct measurement of $S_1$ was made 
using a tensor polarimeter~\cite{POLDER94} at $E_d$=400 MeV~\cite{Furget95}. 
The limited coverage of the experimental setup, however, 
confined the measurement to be done 
only for the $1^+$ (12.71 MeV) state in \nuc{12}{C} at 4$^{\circ}$. 
Using a vector polarimeter~\cite{POMME90} 
the $(\vec{d},\vec{d}')$ measurements were performed 
on a wide range of nuclei~\cite{Morlet90,Johnson95,Djalali99}.  
The vector quantity $S_d^y=\frac43+\frac23A_{yy}-2K_{y}^{y'}$ 
was used as the signal for the isoscalar spin states. 
This quantity, however, is an approximate one, 
which coincides with $S_1$ only when $A_{yy}=P^{y'y'}$ and $S_2$=0. 
In further pursuing a systematic study of the isoscalar spin-flip transitions 
it is therefore important to develop an experimental technique 
to measure $S_1$ over an extended range of the excitation energy. 
Furthermore it is fascinating to see 
if the intriguing existence of $\Delta S$=2 states, 
such as the proposed double GT state~\cite{Vogel88,Auerbach89}, 
could be revealed through $S_2$ as the probe. 

In this letter 
we report on a measurement of the deuteron single and double SPFs 
using a focal plane deuteron polarimeter 
specifically designed to measure both vector and tensor polarization 
components of deuterons. 
The \nuc{12}{C} target was chosen 
as it provides a typical isoscalar spin-flip 1$^+$ state at 12.71 MeV 
along with non-spin-flip collective states, 
therefore affording a study of $\Delta S$-dependent features 
of the deuteron SFPs.  

The experiment was performed at RIKEN accelerator research facility (RARF). 
The polarized deuteron beam from the Ring Cyclotron 
with an energy of 270 MeV and an average current of 10 nA 
was led onto a \nuc{12}{C} target with a thickness of 87.2 mg/cm$^2$. 
The vector and tensor polarizations 
with maximum values of $(p_y, p_{yy})$=(0,0), (0,-2), (2/3,0) and (-1/3,1) 
were used, 
the magnitudes 
were measured using the $d$+$p$ elastic scattering 
at 270 MeV~\cite{Sakamoto96} 
to be 60-70 \% of the ideal values. 
Scattered deuterons were detected 
using the SMART spectrometer~\cite{Ichihara94} 
consisting of quadrupoles (Q) and horizontally bending dipoles (D) 
in the QQDQD configuration. 
It was instrumented with a multiwire drift chamber (MWDC) 
backed by two plastic scintillation detectors 
for track reconstruction and triggering. 
It had a solid angle of 5.0 msr and a momentum acceptance of 4\%. 
The polarizations of outgoing deuterons 
were measured in a deuteron polarimeter DPOL installed in the focal plane. 
Data were acquired at a spectrometer setting of 5$^{\circ}$. 

Measurements of both vector and tensor components 
of the outgoing deuterons 
are crucial in extracting $S_1$ and $S_2$. 
This was realized by utilizing, 
as the analyzer reactions, 
the $\vec{d}+$C elastic scattering 
and the \nuc{1}{H}$(\vec{d},2p)$ charge exchange reaction, 
which show large angular asymmetries 
depending, respectively, on the vector and tensor components. 
The deuterons from the primary scattering 
were incident on the analyzer target 
consisting of a 2.5-cm thick CH$_2$ block 
bracketed by the trigger counters. 
A counter hodoscope consisting of two layers 
of plastic scintillation counter array HOD and CM 
($\sim 2 \times$ 2 m$^2$) was located 4 m behind the analyzer target. 
It was used to detect 
deuterons from the $\vec{d}+$C elastic scattering 
and pairs of protons produced 
through the \nuc{1}{H}$(\vec{d},2p)$ reaction 
with small relative momenta. 
An iron absorber between HOD and CM 
was used to discriminate between deuterons and protons. 
In off-line analysis the analyzer reactions 
were separated from other parasitic ones, 
such as the $\vec{d}$+$p$ and \nuc{12}{C}$(\vec{d},2p)$ reactions, 
using cuts on missing-mass spectra. 
The calibration of DPOL was done 
in the energy range between 230 and 270 MeV~\cite{Sishida98}. 

The analysis of the polarimeter 
has been proceeded in the same way 
as that in the calibration analysis. 
The cross section for spin-1 deuterons 
is given by 
\begin{eqnarray}
\sigma(\theta,\phi)=\sigma_0(\theta)
[
1&+&it_{11}T_{11}(\theta)\cos(\phi)+t_{20}T_{20}(\theta) \nonumber \\
&+&2t_{21}T_{21}(\theta)\cos(\phi)+2t_{22}T_{22}(\theta)\cos(2\phi)
], 
\end{eqnarray}
where 
$\sigma_0$ is the cross section for the unpolarized deuterons, 
$T_{kq}$ the analyzing powers, 
$t_{kq}$ the polarization components of the deuterons, 
$\theta$ and $\phi$ the polar and azimuthal scattering angles. 
For each angular and excitation energy bin in the primary scattering, 
$\sigma_0$ and $T_{kq}$ were calculated 
by interpolating the calibration results 
with respect to the incident deuteron energy. 
Possible differences in beam geometry 
between the calibration and this experiment 
could result in different detection efficiencies, 
which is implicit in $\sigma_0$. 
Corrections for the geometrical effects were made 
by using a Monte Carlo simulation~\cite{Ysatou01a}. 
The polarization components $t_{kq}$ 
were extracted through a $\chi^2$ minimization procedure 
using the measured angular distributions of the analyzer reactions. 
After a small correction for spin precession 
inside the spectrometer ($\sim 9.8^{\circ}$), 
the polarization observables were deduced by using the relation 
given in Ref.~\cite{Ohlsen72}. 

Since the polarization transfer coefficients were extracted 
using differences of cross sections in the secondary scattering 
obtained with different polarization modes, 
they were unaffected by the uncertainties in $\sigma_0$. 
On the other hand, the polarizing powers $P^{y'}$ and $P^{y'y'}$ 
were extracted from double scattering cross sections 
obtained with an unpolarized beam. 
The systematic uncertainties in $P^{y'}$ and $P^{y'y'}$ 
are estimated from measurements of elastically scattered deuterons 
to be less than 0.12 and 0.20, respectively. 
The uncertainty for $P^{y'y'}$ 
contributed less than 0.02 to the uncertainties in $S_1$ and $S_2$. 

\begin{figure}[p]
\includegraphics[angle=-90,width=12cm,clip]{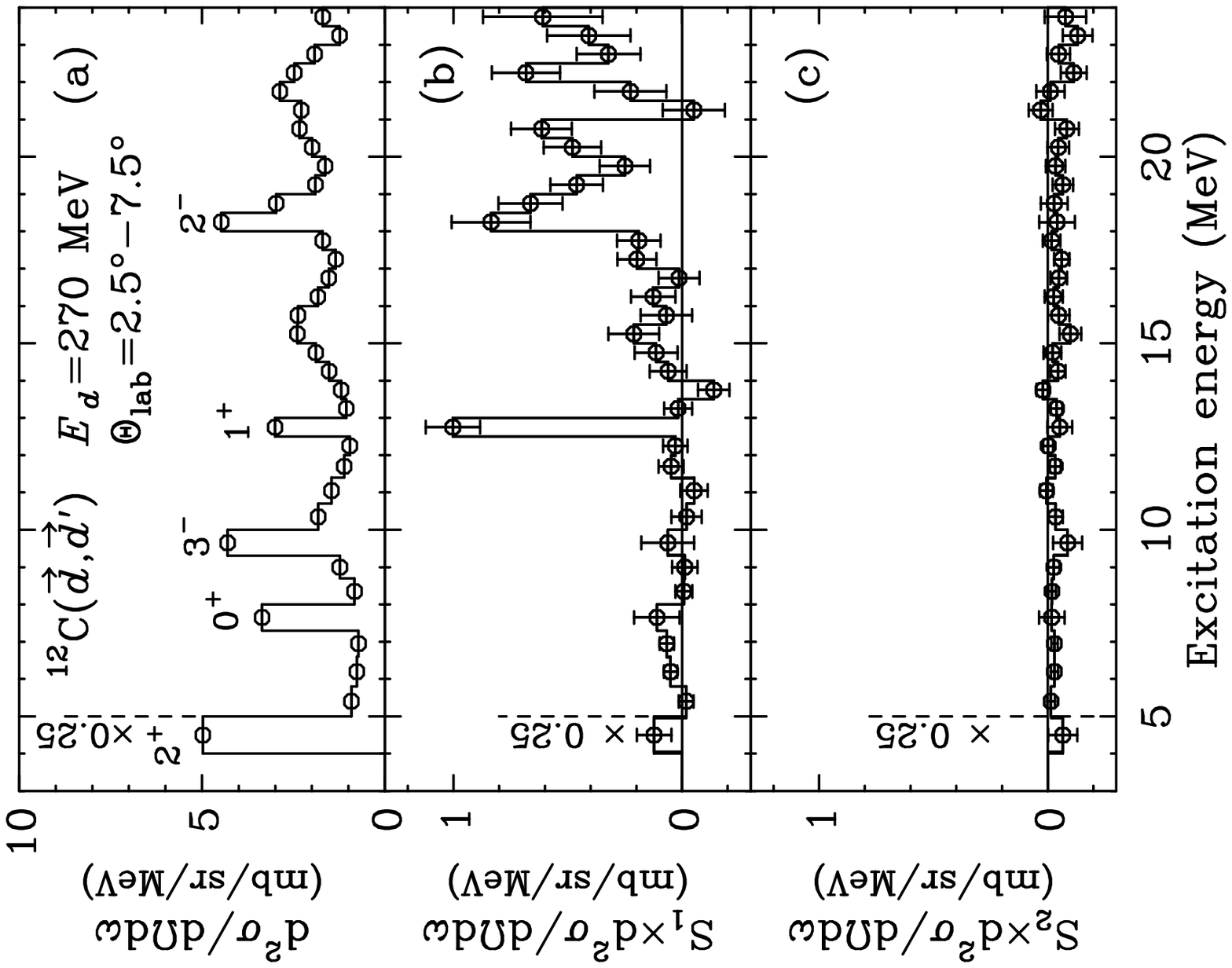}
\caption{Excitation energy spectra 
for the \nuc{12}{C}$(d,d')$ reaction at $E_d$ = 270 MeV 
integrated over $\Theta_L$ = 2.5$^{\circ}$--7.5$^{\circ}$. 
(a) An excitation energy spectrum. 
(b) The spectrum multiplied by $S_1$. 
(c) The spectrum multiplied by $S_2$. }
\label{fig:12C_spectra}
\end{figure}

Figure~\ref{fig:12C_spectra} (a) 
shows the double differential cross sections 
as a function of the $^{12}{\rm C}$ excitation energy. 
The well known isoscalar 1$^+$ (12.71 MeV) state 
is clearly excited along with the other natural parity 
2$^+$ (4.44 MeV), 0$^+$ (7.65 MeV) and 3$^-$ (9.64 MeV) states. 
Isovector states such as the 1$^+$ state at 15.11 MeV 
are absent due to the isospin selection rule. 
Figure~\ref{fig:12C_spectra} (b) 
shows the cross section multiplied by $S_1$. 
The error bars represent the statistical and systematic errors 
added in quadrature. 
We see that the 1$^+$ $\Delta S$=1 transition is strongly enhanced, 
while for the other $\Delta S$=0 natural parity transitions 
the strengths are suppressed. 
This $\Delta S$ dependence of $S_1$ indicates 
that the measurement of $S_1$ in the $(\vec{d},\vec{d}')$ reaction 
shows promise of becoming a useful tool in search for the isoscalar 
spin strength in heavier nuclei. 
The state at 18.3 MeV 
has been assigned as ($J^{\pi}, T$) = (2$^-$, 0) 
from the $S_{nn}$ measurement in
$(\vec{p},\vec{p}\,')$~\cite{Jones83}. 
The present result, 
which exhibits a large $S_1$ strength for this state, 
shows consistency with its identification 
as an isoscalar spin-flip transition. 
Figure~\ref{fig:12C_spectra} (c) plots the cross section multiplied by $S_2$. 
The values are consistent with zero over the measured 
excitation energy region, and 
no clear indication of the $\Delta S$=2 states 
has been obtained from the present measurement. 

\begin{figure}[p]
\includegraphics[width=12cm,clip]{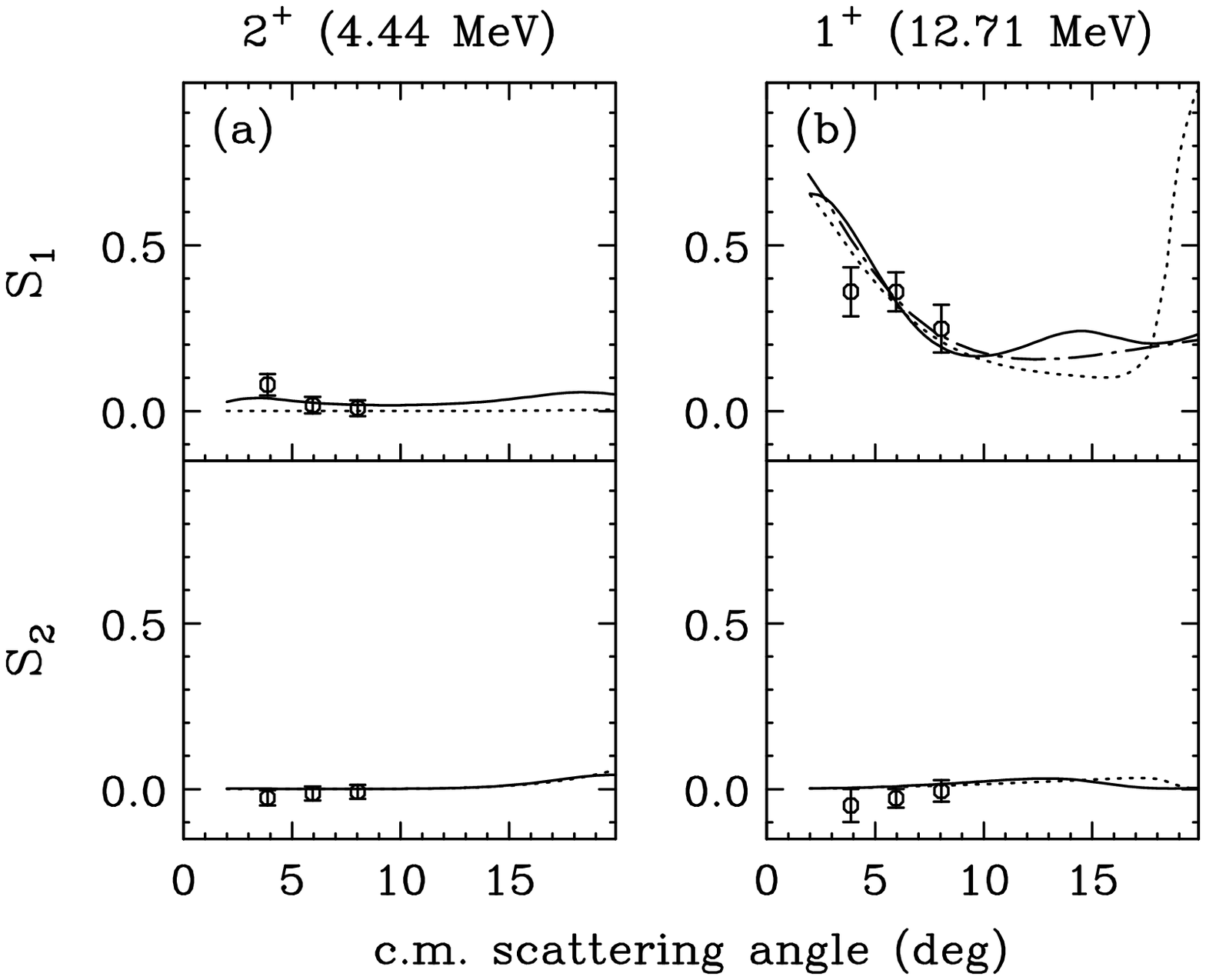}
\caption{The angular distributions of $S_1$ and $S_2$ 
for (a) the 2$^+$ (4.44 MeV) state 
and (b) the 1$^+$ (12.71 MeV) state. 
The solid (dashed) line is the DWIA (PWIA) calculations. 
The dot-dashed line for $S_1$ for the 1$^+$ state 
is an estimate using the IA amplitudes for $LSJ$=(011). }
\label{fig:sfp_f2p_f1p_II}
\end{figure}

Figures~\ref{fig:sfp_f2p_f1p_II} (a) and~\ref{fig:sfp_f2p_f1p_II} (b) 
show the angular distributions of the SPFs for the $2^+$ and $1^+$ states, 
respectively.
The $S_1$ values for the $2^+$ state are close to zero, 
while those for the $1^+$ state 
are large at small scattering angles.  
The $S_2$ data points are consistent with zero for both transitions. 
The results of the microscopic DWIA calculations 
are shown as solid lines in Fig.~\ref{fig:sfp_f2p_f1p_II}. 
Details of the calculation formalism can be found in Ref.~\cite{Wiele95}. 
The one-body transition densities 
were obtained using the $1p$ shell wave functions 
of Cohen and Kurath~\cite{Cohen65}. 
The optical potential 
was determined by fitting the elastic scattering data. 
The projectile-nucleon effective interaction 
was taken to be the on-shell deuteron-nucleon $t$-matrix 
given by~\cite{Tsuzuki94} 
\begin{eqnarray}
t_{dN}(q)=\alpha 
 &+&\beta S_n+\gamma \sigma_n+\delta S_n \sigma_n +\epsilon S_q \sigma_q 
  + \zeta S_p \sigma_p +\eta Q_{qq} +\xi Q_{pp} \nonumber \\
 &+&\kappa Q_{qq} \sigma_n +\lambda Q_{pp} \sigma_n +\mu Q_{nq} \sigma_q  
  + \nu Q_{np} \sigma_p. 
\label{eqn:tdn}
\end{eqnarray} 
The amplitudes $\alpha$ through $\nu$ 
were determined 
from the three-nucleon Faddeev calculations~\cite{Glockle96}. 
The PWIA results obtained by turning off the optical potential 
are also shown as dashed lines in Fig.~\ref{fig:sfp_f2p_f1p_II}. 

As shown in Fig.~\ref{fig:sfp_f2p_f1p_II} 
the overall features of the experimental angular distributions of $S_1$ 
for the $2^+$ and $1^+$ states 
are reasonably described by the present calculations. 
Small but finite DWIA values of $S_1$ for the $2^+$ state 
is due to the spin-orbit distortion in the optical potential. 
The dot-dashed line of $S_1$ for the $1^+$ state 
represents an IA prediction 
for transferred orbital, spin and total angular momenta of $LSJ$ = (011) 
calculated by the relation~\cite{Tsuzuki94}: 
\begin{equation}
S_1=\frac{2|\epsilon|^2+2|\zeta|^2+2|\mu|^2+2|\nu|^2}
{3|\gamma|^2+2|\delta|^2+2|\epsilon|^2+2|\zeta|^2+|\kappa|^2 
+|\lambda|^2-|\kappa+\lambda|^2/3+|\mu|^2+|\nu|^2}. 
\label{eqn:s1_IA}
\end{equation}
The close agreement 
between the DWIA and this simple IA estimate below $10^{\circ}$ 
indicates that $S_1$ for the $1^+$ state in this angular region
is relatively insensitive to distortion effects 
as well as to details of the transition density, 
instead it is mainly determined by the relative strengths 
of various parts of the IA amplitudes. 
This characteristic behavior of $S_1$ at forward angles 
is similar to that reported for $S_{nn}$ in $(p,p')$~\cite{Morris82}. 
The calculated $S_2$ values for both $2^+$ and $1^+$ states 
are consistent with zero below 10$^{\circ}$. 
Since 
the effective interaction used in the present calculations 
adequately takes into account correlation effects 
among the projectile-nucleon system, 
the calculated results suggest that 
any possible non-zero values for $S_2$ at forward angles 
would have to be explained by reaction mechanisms 
which are not included in the present calculation, 
such as two-step processes involving spin-flip 
of two target nucleons, 
and tensor terms in the optical potential. 

In summary, 
we have succeeded in measuring the single and double spin-flip 
probabilities 
through direct observation of polarization transfer 
in inelastic deuteron scattering on $^{12}{\rm C}$ at $E_d$ = 270 MeV. 
The focal plane polarimeter, 
which utilized the $\vec{d}+$C elastic scattering 
and the \nuc{1}{H}$(\vec{d},2p)$ charge exchange reaction, 
allowed all the polarization components of the scattered deuterons 
to be measured simultaneously 
over a wide excitation energy range (4 and 24 MeV). 
The $S_1$ value is large for the spin-flip $1^+$ (12.71 MeV) state, 
while it is close to zero for other non-spin-flip states, 
such as the first $2^+$ (4.44 MeV) state. 
The $S_2$ values are consistent with zero 
for all the measured excitation energy range. 
The overall trends of the SFP angular distributions 
for the $1^+$ and $2^+$ states 
are well described by the microscopic DWIA calculations. 
The demonstrated feasibility of measuring the deuteron SFPs 
as well as 
the capability of the DWIA theory to reproduce the data 
will add a new and an alternative probe of nuclear structure. 
Further experiments 
are planed in search for the $\Delta S$=1 and 2 transitions 
in higher excitation energy region in \nuc{12}C and in other nuclei. 

We thank Dr.\ H.\ Kamada for providing the deuteron-nucleon amplitudes. 
We are grateful to the staff of RARF 
for their invaluable assistance during experiment. 
This work was supported 
by 
the Special Postdoctoral Researchers Program at RIKEN.

\end{document}